\documentclass[sigconf, nonacm]{acmart}

\usepackage{siunitx}
\usepackage[most]{tcolorbox}
\usepackage{tabularx}
\begin{document}

\title{AutoGrad: Automated Grading Software for Mobile Game Assignments in SuaCode Courses}

\author{Prince Steven Annor}
\affiliation{\institution{New York University Abu Dhabi}}
\affiliation{\institution{Nsesa Foundation}}
\email{princeannor@gmail.com}

\author{Samuel Boateng}
\affiliation{\institution{Nsesa Foundation}}
\email{kaybeta500@gmail.com}

\author{Edwin Pelpuo Kayang}
\affiliation{%
  \institution{University of Ghana}}
\affiliation{\institution{Nsesa Foundation}}
\email{pelpuo2000@gmail.com}

\author{George Boateng}
\affiliation{%
  \institution{ETH Z{\"u}rich}}
\affiliation{
  \institution{Nsesa Foundation}}
\email{gboateng@ethz.ch}

\renewcommand{\shortauthors}{Annor et al.}

\begin{abstract}
Automatic grading systems have been in existence since the turn of the half-century. Several systems have been developed in the literature with either static analysis and dynamic analysis or a hybrid of both methodologies for computer science courses. This paper presents AutoGrad, a novel portable cross-platform automatic grading system for graphical Processing programs developed on Android smartphones during an online course. AutoGrad uses Processing, which is used in the emerging Interactive Media Arts, and pioneers grading systems utilized outside the sciences to assist tuition in the Arts. It also represents the first system built and tested in an African context across over thirty-five countries across the continent. This paper first explores the design and implementation of AutoGrad. AutoGrad employs APIs to download the assignments from the course platform, performs static and dynamic analysis on the assignment to evaluate the graphical output of the program, and returns the grade and feedback to the student. It then evaluates AutoGrad by analyzing data collected from the two online cohorts of 1000+ students of our SuaCode smartphone-based course. From the analysis and students' feedback, AutoGrad is shown to be adequate for automatic assessment, feedback provision to students, and easy integration for both cloud and standalone usage by reducing the time and effort required in grading the 4 assignments required to complete the course.
\end{abstract}



\maketitle

\section{Introduction}
The growth of online learning has skyrocketed with millions of users joining eLearning platforms due to the added flexibility and low costs of course delivery \cite{RN327}. Conventional universities are also adopting online learning as separate distance learning tracks or as supplements to their lessons to scale and increase engagement outside the physical lecture hall. The importance of online learning has become even more apparent now with the ongoing COVID-19 pandemic which has caused several schools all over the world to suspend in-person teaching and to deliver their courses online in a bid to implement physical distancing measures. 

Students across Africa are likely to feel the impact of this switch especially in the area of programming education with the existing limited access to computers which tend to be in schools and not in homes. However, there is a rapid expansion of smartphone ownership and usage in Africa. Research from Ovum shows that smartphone ownership will hit 929.9 million smartphones in Africa this year \cite{matinde2017}. Hence, there is a unique opportunity to use smartphones for online computer science education in Africa.

Therefore, in 2017, a smartphone-based coding course was developed in Ghana for an annual summer bootcamp, which was later developed into an online course in 2018 and scaled across the African continent to over 700 students in 2019 \cite{boateng2018, boateng2019,boateng2021} and 2000+ students in 2020 \cite{boateng2020}. The smartphone-based course used Processing with the Android Processing Development Environment (APDE) \cite{apde} to introduce students to the basic concepts in Processing and graphical programming (Lesson 1), Variables (Lesson 2), Conditionals (Lesson 3) and Functions (Lesson 4) and resulted in the building of a Pong game \cite{boateng2019, apde}. The 72\% of students that completed in the 2019 cohort were therefore able to use APDE on their smartphones to produce a functional video game from Processing built into an Android APK at the end of the course.

Processing is a Java-based graphical language used in the visual arts for learning how to code and for prototyping several installations, simulations and games \cite{processing}. Processing is open source and cross platform (Windows, Mac OS X, GNU/Linux, Android, and ARM), and it has OpenGL integration for 2D and 3D acceleration \cite{processing}. By 2020, the artists, students, researchers, designers and hobbyists that use Processing are in the tens of thousands, and institutions like New York University are using Processing in their computer science and Interactive Media curricula \cite{processing}. The Android Processing Development Environment is an Android application that provides an IDE for developing Processing programs and building them into wallpapers, android, watch and VR applications.

Processing also makes the development of mobile applications very simple and interactive and data from the SuaCode course has shown that students enjoyed coding on their smartphones (4.5±0.78 on a scale from 1 to 5). However, like other mobile application courses, assessment is nontrivial \cite{RN328}, so institutions employ teaching assistants to grade in such courses. This approach is not scalable especially in the context of online courses such as SuaCode with thousands of learners. In addition, although research and development has been done to automate assessment of some graphical and mobile application assignments, none has been done for emerging languages geared towards the emerging Interactive Arts such as the Processing language \cite{RN330, RN322, RN331}. Moreover, none are fully integrated to an online course system and none perform syntax, semantic, style in addition to dynamic analysis.

The importance of developing a grading system as the smartphone-based course scaled to hundreds across Africa became even more apparent since each lesson had an assignment in addition to the final project. Consequently, we developed such a system — AutoGrad. This paper presents AutoGrad, a novel portable cross-platform automatic grading system for graphical and interactive programs written in the Processing programming language on Android smartphones during SuaCode courses. Firstly, we present the design and implementation of AutoGrad. AutoGrad employs APIs to download the assignments from the course platform, performs static and dynamic analysis on the assignment to evaluate the graphical output of the program, and returns the grade and feedback to the student. Secondly, we evaluate AutoGrad by analyzing data collected from the online courses of students that used AutoGrad. From the analysis, AutoGrad is shown to be effective for automatic assessment, feedback provision to students and easy integration for both cloud and standalone usage. Given the usefulness of a Processing-based smartphone course, this work will eventually enable the scaling of coding education to several people across the African continent.

\section{Background and Related Work}
Automatic grading systems have been in existence since the turn of the half-century. In addition, several developments have been made for the facilitation of programming pedagogy \cite{RN340, RN339, RN338}, and automatic assessment analysis with either static analysis and dynamic analysis or a hybrid of both methodologies \cite{RN333, RN330, RN322, RN334, RN328, RN332, RN327, RN336, RN335, RN331, david_liv}. Maicus et al. employed containers for automatic grading for distributed algorithms courses and evaluated the submissions by examining the data streams between components \cite{RN341}. However, our work is concerned with standalone mobile applications (games) developed with Processing that are graphical in nature and have user interactivity as a key component of the program. 

Specific work of interest includes dynamic analysis methodologies that leverage virtual machines and containers for running student submissions and evaluating graphical output. Wünsche et al. developed an automatic grading system that used a Moodle plug-in for assessing the student code in a virtual machine \cite{RN331}. Bruzual et al. developed a mobile app testing framework that used Docker containers to assess the student code by using unit tests on the Android application package (APK) \cite{RN322}. Our work can be deployed both in the cloud, with virtual machines and containers for resource scaling in massive open online courses, and on standalone systems such as computers and smartphones. Our system could be provided directly to the student for self-evaluation of the work, but that deviates from the pedagogy style employed in the course used to evaluate the system. Lastly, this work can work with the actual code before it is compiled and packaged to the APK or other formats that Processing can compile to. 

This work follows the automatic assessment style of English \cite{jewl}, where a set of Java packages, namely JEWL, are used both for teaching and for assessing visual elements dynamically. Similarly, another Java package, namely Objectdraw, was used by Thornton et al. for supporting student-written tests \cite{thornton}. Gray et al. also present an introspective approach for dynamic assessment of Java GUI programs by using a parser written in Java to directly extract information from the visual objects  \cite{gray}. Our work presents a more holistic grading approach, however, where function calls, variable declarations, and acceptable coding standards like commenting and indentation, etc are detected and graded in addition to the visual assessment. Our work also uses Processing, which is used in the emerging Interactive Media Arts, and pioneers grading systems outside the sciences and engineering to assist tuition in the Arts. Lastly, our work represents the first system built and tested in an African context across over thirty-five countries across the continent.




\begin{figure}
    \centering
    \includegraphics[width=0.7\linewidth]{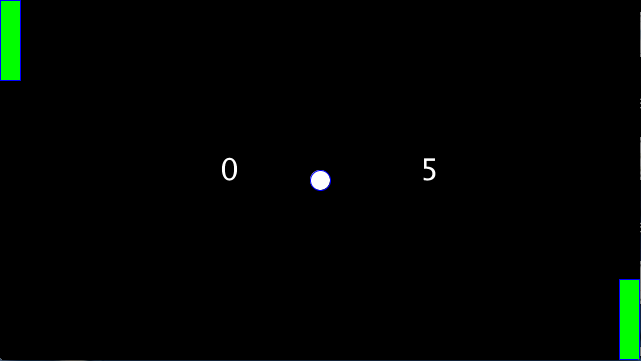}
    \caption{Expected output for Assignment 1}
    \Description{Figure showing code output for SuaCode assignment 1}
    \label{fig:Assignment 1}
\end{figure}

\section{Overview of AutoGrad}
\begin{figure}
    \centering
    \includegraphics[width=\linewidth]{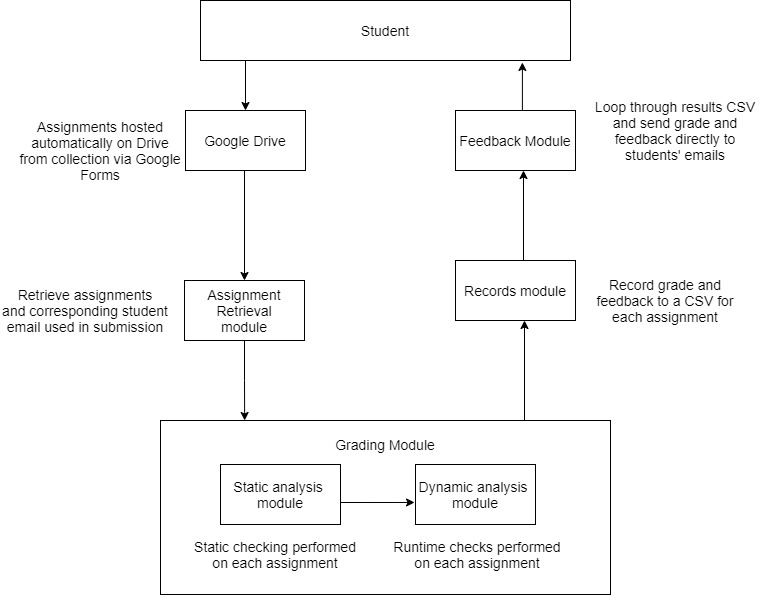}
    \caption{System design of AutoGrad}
    \Description{}
    \label{fig:autograd}
\end{figure}

AutoGrad is a novel portable cross-platform automatic grading system for graphical and interactive programs written in the Processing programming language on Android smartphones for SuaCode courses. AutoGrad employs APIs to download the assignments from the course platform containing the assignment submissions, performs static and dynamic analysis on the assignment to evaluate the graphical and interactive output of the program, and returns the grade and feedback to the student (Figure \ref{fig:autograd}). Details are provided in the next sections.

Processing language is Java-based programming language that contains libraries that make it easy to write interactive, graphical programs such as games and simulations \cite{processing} making it easy to introduce novices to coding in an interesting and fun way. For example, the Processing language has out-of-the-box methods for creating and formatting graphical elements such as lines and shapes (e.g. line(), ellipse(), rect(), text() for adding a line, ellipse, rectangle and text respectively). So to check if an ellipse has been drawn at the center, the task becomes finding if there is a call to the "ellipse()" function and checking the parameters to ensure it is at the center. This reduced our efforts to check certain assignment specifications and also reduces variability when writing test cases to identify where certain graphical elements have been created.

We wrote the AutoGrad system in Python and Processing. The Python code handles the retrieval of assignments and returning the assignment grade and feedback to students. Processing handles the grading module so it adequately implements the checks for the Processing code like we described previously.

We have deployed AutoGrad in two different cohorts of SuaCode for use in grading scripts of 1000+ students across Africa \cite{boateng2021,boateng2020} with over 3,000 code files graded. Those deployments involved AutoGrad being ran as a software on a computer. Additionally, for the most recent cohort, we provided students the opportunity to submit complaints when they felt their assignments were incorrectly graded. Doing this gave us the opportunity to be fair to all students, address cases where AutoGrad was not adequate and also to improve the software.

\begin{table}[]
\centering
\caption{Detail on the assignments in the SuaCode course}
\resizebox{0.45\textwidth}{!}{%
\begin{tabular}{|p{1.4cm}|p{2.5cm}|p{4.2cm}|}
\hline
\label{tab:Assignment_description}
Assignment Number & Assignment Title & Description \\ \hline
1                 & Make Pong Interface                 & The assignment involves the use of in built processing functions to design the interface of the pong game. The interface is simple, containing two rectangles to serve as paddles, a circle to serve as a ball and two text elements to represent scores.             \\ \hline
2                 & Move Ball                 & Students are to simulate ball movement by using operations on variables to change the ball’s location after each frame. The additional task is to replace all used values with variables thus removing magic numbers            \\ \hline
3                 & Bounce Ball                 & The goal of this assignment is to make use of conditional statements to reverse the direction of the ball when it hits the ends of the screen            \\ \hline
4                 & Move Paddles                 & This assignment requires students to employ knowledge on functions to move the paddles, make the ball bounce off the paddles and separate other parts of the code into functions such as displayScores(), displayBall() and displayPaddles()            \\ \hline
\end{tabular}
}
\end{table}

\subsection{Structure of Course Assignments}
The curriculum covers four lessons: Basic Concepts in Processing (Lesson 1), Variables (Lesson 2), Conditionals (Lesson 3) and Functions (Lesson 4),  and results in the building of a Pong game  (Figure \ref{fig:Assignment 1}) \cite{boateng2018, boateng2019}. This game has 2 paddles, one for each player and a ball. Once the ball starts moving, each player has one goal — to prevent the ball from exiting the vertical wall on their side using the paddles. If the latter happens, the opponent’s score increases. Lesson 1 introduces students to Processing and some basic concepts in graphical programming. Lesson 2 then builds upon lesson 1 and introduces variables and standard coding practices. Lesson 3 introduces conditionals, specifically if-else statements, and lesson 4 introduces functions and best modular coding practices. Each lesson has a corresponding assignment that incrementally builds a component of a Pong game.
  
Assignment 1 entails building the interface of the pong game, assignment 2 makes the ball move by incorporating variables to store state, assignment 3 makes the ball bounce off the game walls by using conditionals, and assignment 4 gets students to write functions to move the paddles and put it all together (Table \ref{tab:Assignment_description}). Here is an example of the instructions and specifications that students get for assignment 1.

Using your Assignment1, write code to draw a Pong game interface like in the picture above with the following specs:
\begin{itemize}
\item Two paddles, one at the top left end exactly and one at the bottom right end exactly pick your own width and height but they should be the same for both paddles 
\item Ball at center 
\item Pick your circle’s width and height
\item Write 2 arbitrary numbers, one at the left side and the other at the right side of the screen, representing the left and right players’ scores respectively
Set the size of the text
\item Use different colors for the interior of the paddles, ball and window background
\item Color should be the same for both paddles
\item Set one color for the outlines of all shapes
\item Use setup() and draw() to organize your code
\item Add comments to your code
\item Your code looks cleaner when you group lines of code together like grouping all code of the ball together, and then the code for the paddles together, then code for the scores together, etc.
\item Make sure to indent your code properly
 
\end{itemize}
\subsection{Grading Criteria}
Using the specifications from the assignment given to students as criteria, instructors draw up a grading criteria document to convert assignment specifications to a checklist of values or outputs the grading system should look out for. A typical example of grading criteria for assignment 1 is as follows:

\begin{enumerate}
\item Screen size is the whole screen: fullScreen()
\item Left paddle at the top left corner of the screen: x = 0, y = 0
\item Right paddle at the bottom right of the screen: x= width-paddleWidth, y = height-paddleHeight
\item Both paddles have the same width and height: leftPaddleWidth = rightPaddleWidth, leftPaddleHeight = rightPaddleHeight
\item Ball at center and is a circle: x = width/2, y = height/2, w = h
\item Left player’s score on left side of screen: x < width/2
\item Right player’s score on right side of screen: x > width/2
\item Size of the text is set: textSize()
\item Interior of both paddles have the same color
\item Colors for the interior of the paddles different from ball and also from screen background 
\item Outlines of all shapes have one color: stroke()
\item Code organized with setup() and draw() 
\item Code well commented: 2 backslashes come after semicolon in each line of code or 2 backslashes above blocks of code + description about that line of code
\item Code indented properly:  1 tab at start of all code in setup() and draw()
\item Code looks clean: each line of code on a separate line, all code of the ball grouped together, and then the code for the paddles together, then code for the scores together, etc.
\end{enumerate}

\subsection{Modular Test Methods for Assignments}
“Test methods” which are essentially modular functions are created for the various criteria. Test methods are written by course creators. Each test method validates students’ code submissions against the grading criteria and can then deducts points when a test fails. For example, the checkTabs() test method checks if the student’s script uses proper indentation, while checkScores checks if the student created text to represent scores and at the appropriate placement on the screen according to the assignment.
Examples of test methods for the SuaCode course, written by course creators for Assignment 1 are as follows:


\begin{table}[]
\caption{Modular functions used in grading Assignments 1 to 4 with short explanations of what they do.}
\label{tab:Assignment_1_modular}
\resizebox{0.45\textwidth}{!}{%
\begin{tabular}{|p{0.9em}|p{9.5em}|p{14.4em}|}
\hline
No. & Modular function name & Functionality \\ \hline
1 & checkTabs & checks if the right amount of indentation has been applied on each line in each block of code throughout the source file \\ \hline
2 & checkStatementsPerLine & checks if statements are written 1 per line according to source code formatting guidelines, \\ \hline
3 & checkComments & checks if student has written at least enough comments as per assignment guidelines (30 percent of code should be commented) \\ \hline
4 & checkBackground & checks if the background colour is as specified in assignment \\ \hline
5 & checkFills & checks if shapes are filled with the expected colours \\ \hline
6 & checkStrokes & checks if shape outlines have the expected colours \\ \hline
7 & checkEllipses & checks if ellipses exist and are drawn with the expected dimensions \\ \hline
8 & checkRects & checks if rectangles and squares exist and are drawn with the right dimensions. \\ \hline
9 & checkScores & checks if the scores (text) exist in the code and have the right values \\ \hline
10 & checkMovingBall &  checks if the ball moves when code is run \\ \hline
11 & checkGameOn & checks if the game starts when the mouse is pressed \\ \hline
12 & checkWallsBounce Top & checks if the ball bounces off the top of the screen when the game is running \\ \hline
13 & checkWallsBounce Bottom & checks if the ball bounces off the bottom of the screen when the game is running \\ \hline
14 & checkLeftWall & checks if the ball exits the left of the screen and increments the right score when the game is running \\ \hline
15 & checkRightWall & checks if the ball exits the right of the screen and increments the left score when the game is running \\ \hline
16 & checkCreatedFunctions Exist & checks if the functions named in the assignment are created \\ \hline
17 & checkMoveLeftPaddle & checks if the left paddle moves along with the mouse \\ \hline
18 & checkMoveRightPaddle & checks if the right paddle moves along with the mouse \\ \hline
19 & checkBounceLeftPaddle & checks if the ball bounces off the left paddle when it collides with it \\ \hline
20 & checkBounceRightPaddle & checks if the ball bounces off the right paddle when it collides with it \\ \hline
\end{tabular}
}
\end{table}

\subsection{Static and Dynamic Code Analysis}
The grading module runs both static and dynamic code analysis and compares the student’s code with the specifications for the assignments. 

\subsubsection{Static Analysis}
One key benefit of using the Processing development environment is the availability of functions for making graphical elements such as the ellipse() function for drawing ellipses and circles, the rect() function for drawing rectangular figures among others. This simplifies the effort of checking what students have done to checking what has been supplied as parameters to these functions. The main technique used in static analysis is text matching using regular expressions. Various important pieces of information can be verified using this approach without manual inspection from instructors.

With static analysis, we first parsed the student’s code file and employed regular expressions checks. An example of static analysis is as follows. In assignment one, students have to draw a ball at the center of the screen. AutoGrad in this case checks if a circle has been drawn via searching for the function call for a circle and where it was drawn based on its parameters.

\subsubsection{Dynamic analysis of scripts}
When various criteria could not be easily assessed just by looking at how the code was written, we employed both static and dynamic analysis to help us see what is happening. This proved particularly useful in the latter assignments, Assignment 3 and 4 where it was necessary to track changes to the value of a variable with time during execution. For example, in order to see whether a ball created in a student’s script actually moves in a certain direction, the variables that store the position of the variable need to be identified and then observed for at least more than two frames. 

With dynamic analysis, the student’s code gets wrapped into a class definition, an object gets instantiated to run the code file, and then various checks are performed by manipulating various parameters and checking how other parameters change. An example of dynamic analysis is checking if, for example, the ball bounces off the top and bottom walls correctly. To do this, we run the student’s code, make the ball move towards the top and bottom walls, and check if various parameters such as the ball’s movement direction changes as expected.

The Processing development environment requires that code be written in two main functions, setup and draw. One time initialization go into setup and instructions that need to be executed repeatedly go into draw. Before calling test methods to grade a script for assignments that need dynamic analysis, we transfer the setup and draw functions into a custom class along with all variable declarations and other functions defined by the student, by calling a preliminary script.

The actual grading script is called after the preliminary script by then which the grading script has a class to work with, with all the information from the student’s script contained within it as attributes and methods. Then we can use Java Reflections API to be able to dynamically access the methods and properties during run-time. The preliminary script also gives us the opportunity to catch errors in student’s scripts before trying to run them in the grading script to avoid run-time errors.

\subsection{Grades and Feedback from AutoGrad}
AutoGrad provides students with feedback on their assignment code submissions with pointers to missing aspects of the assignment’s specifications (e.g., your code was not indented properly). AutoGrad assigns a total score for that assignment by subtracting points (e.g. 1 or 2) for each of the specifications that are not met from total possible 20 points. Each check has a corresponding feedback to be given if the check is not satisfied. The feedback provided is high-level in order not to fix the problem for the student but give them the chance to relook at their work and address it themselves. This grading summary consisting of scores and feedback is then mailed to the student. Examples of feedback from test methods for the SuaCode course are as follows: 

\begin{itemize}
\item Use at least one stroke() function
\item You may not have called or supplied arguments to the ellipse() function
\item You may not have called or supplied proper arguments to the rect() function
\item You may not have used the background() function
\item You may not have called or supplied proper arguments to the text() function
\item You may not have called or supplied proper arguments to the fill() function
\end{itemize}

\subsection{Complementary Software for the Grading system}
For agility in development, we wrote code in the Python programming language to wrap the core Processing grading system to automate non-grading specific tasks namely:

\begin{itemize}
    \item Downloading assignment code files submitted via Google Forms with the help of the Google Drive Python module
    \item Calling AutoGrad core script in Processing to start grading
    \item Creating emails out of CSV results from AutoGrad core script that were then sent to students inboxes using the Google Mail Python module
\end{itemize}

\section{Evaluation}
We performed different evaluations such as comparing AutoGrad's grading with those of human graders. Additionally, we evaluated students' quantitative and qualitative feedback on their experience with AutoGrad.

\subsection{Comparison between AutoGrad and Human Graders}
We evaluated the test methods to ensure that they will satisfactorily grade student’s scripts. We used test assignment scripts as well as actual student scripts from former cohorts of the SuaCode course to make sure the test methods gave appropriate grades for each script during the design. The results are shown in Tables \ref{tab:Assignment_1_comparison},  \ref{tab:Assignment_2_comparison}, \ref{tab:Assignment_3_comparison}, and \ref{tab:Assignment_4_comparison}. Out of a sample of Assignment 1 scripts of 10 students, AutoGrad gave the same grade as an instructor manually grading the scripts in 8 scripts, computing to 0.7 mean absolute error. Subsequent assignments had more errors compared to assignment 1.

In scripts which had different results, a number of reasons were accountable for the difference, all of which were shortfalls of the AutoGrad software, namely:

\begin{itemize}
    \item  Using multiple term mathematical expressions as parameters which could not be evaluated by static analysis done for that version of the software. Example is ellipse(x+20, 100, 20, 20), where the first parameter has a mathematical expression.
    \item Inline declaration of variables which could not be detected easily using static checking/analysis for that version of the software.
\end{itemize}


\begin{table}[]
\caption{A comparison of grades given by AutoGrad to instructor assigned grades for ten students for Assignment 1.}
\label{tab:Assignment_1_comparison}
\resizebox{0.45\textwidth}{!}{%
\begin{tabular}{|p{1.3em}|p{7.5em}|p{7.4em}|p{8em}|}
\hline
No. & AutoGrad's grade (max 20 marks) & Teacher's grade (max 20 marks)  & Error (e) \\ \hline
1 & 17 & 17 & 0 \\ \hline
2 & 12 & 18 & 6 \\ \hline
3 & 18 & 18 & 0 \\ \hline
4 & 10 & 9 & 1 \\ \hline
5 & 16 & 16 & 0 \\ \hline
6 & 20 & 20 & 1 \\ \hline
7 & 17 & 17 & 0 \\ \hline
8 & 11 & 11 & 0 \\ \hline
9 & 17 & 17 & 0 \\ \hline
10 & 17 & 17 & 0 \\ \hline
 &  &  & Mean absolute error = 0.7 \\ \hline
\end{tabular}
}
\end{table}

\begin{table}[]
\centering
\caption{A comparison of grades given by AutoGrad to instructor assigned grades for ten students for Assignment 2.}
\label{tab:Assignment_2_comparison}
\resizebox{0.45\textwidth}{!}{%
\begin{tabular}{|p{1.3em}|p{7.5em}|p{7.4em}|p{8em}|}
\hline
No. & AutoGrad's grade (max 20 marks) & Teacher's grade (max 20 marks)  & Error (e) \\ \hline
1 & 20 & 17 & 3 \\ \hline
2 & 20 & 16 & 4 \\ \hline
3 & 6 & 8 & 2 \\ \hline
4 & 16 & 12 & 8 \\ \hline
5 & 19 & 17 & 2 \\ \hline
6 & 19 & 17 & 2 \\ \hline
7 & 15 & 12 & 3 \\ \hline
8 & 7 & 8 & 1 \\ \hline
9 & 14 & 11 & 3 \\ \hline
10 & 19 & 12 & 7 \\ \hline
 &  &  & Mean absolute error = 3.5 \\ \hline
\end{tabular}
}
\end{table}

\begin{table}[]
\centering
\caption{A comparison of grades given by AutoGrad to instructor assigned grades for ten students for Assignment 3.}
\label{tab:Assignment_3_comparison}
\resizebox{0.45\textwidth}{!}{%
\begin{tabular}{|p{1.3em}|p{7.5em}|p{7.4em}|p{8em}|}
\hline
No. & AutoGrad's grade (max 20 marks) & Teacher's grade (max 20 marks)  & Error (e) \\ \hline
1 & 14 & 16 & 2 \\ \hline
2 & 16 & 17 & 1 \\ \hline
3 & 3 & 13 & 10 \\ \hline
4 & 14 & 14 & 0 \\ \hline
5 & 19 & 19 & 0 \\ \hline
6 & 19 & 19 & 0 \\ \hline
7 & 18 & 19 & 1 \\ \hline
8 & 18 & 18 & 0 \\ \hline
9 & 3 & 5 & 2 \\ \hline
10 & 18 & 19 & 1 \\ \hline
 &  &  & Mean absolute error = 1.7 \\ \hline
\end{tabular}
}
\end{table}

\begin{table}[]
\centering
\caption{A comparison of grades given by AutoGrad to instructor assigned grades for ten students for Assignment 4.}
\label{tab:Assignment_4_comparison}
\resizebox{0.45\textwidth}{!}{%
\begin{tabular}{|p{1.3em}|p{7.5em}|p{7.4em}|p{8em}|}
\hline
No. & AutoGrad's grade (max 20 marks) & Teacher's grade (max 20 marks)  & Error (e) \\ \hline
1 & 11 & 8 & 3 \\ \hline
2 & 17 & 17 & 0 \\ \hline
3 & 12 & 18 & 6 \\ \hline
4 & 16 & 19 & 3 \\ \hline
5 & 19 & 19 & 0 \\ \hline
6 & 6 & 8 & 2 \\ \hline
7 & 20 & 20 & 0 \\ \hline
8 & 17 & 16 & 1 \\ \hline
9 & 14 & 14 & 0 \\ \hline
10 & 20 & 19 & 1 \\ \hline
 &  &  & Mean absolute error = 1.6 \\ \hline
\end{tabular}
}
\end{table}

\subsection{Students' feedback on Experience with AutoGrad}
We collected quantitative and qualitative feedback from students on their experiences with AutoGrad. For the quantitative feedback, we asked students to respond to the statement "I liked the feedback from AutoGrad" on a 5-point Likert scale from strongly disagree to strongly agree. Out of the 457 students of the 2020 cohort that completed the course and responded, 75.9\% of students agreed or strongly agreed with the statement showing that several students found AutoGrad's feedback useful with some room for improvement with mean = 4 out of 5. For the qualitative feedback, we asked for suggestions for improving the AutoGrad's feedback. Most of the responses  asked for detailed explanations for the feedback AutoGrad gives and that it also point outs where exactly in their code they went wrong. Other students pointed out the need to improve the accuracy of AutoGrad's grading.


 
\subsection{Implications of results}
The AutoGrad system is currently fairly able to handle assignments from students which do not deviate significantly from the instructions given for the assignment. However it implies that extra care needs to be taken when designing assignments and laying out steps for students to follow such that they can be easily handled by the AutoGrad system. Additionally, we should consider providing more detailed feedback to students without also solving the problem for them.

\section{Limitations and Future Work}
The current version of AutoGrad has been deployed for only one course in two cohorts so far (1,000+ students) and we plan to deploy in other courses to further ensure that the system is robust for a wide range of courses. Currently, course creators are responsible for writing test methods which are then used to automate the grading of new courses. Our major next step and future work will be to make easy the effort required to create these checks and test cases for new courses written in Processing. We have actually already began this process by packaging these checks in modular functions which can be imported and modified (e.g. checkIndentation which checks for proper indentation, checkBall, which checks for the placement of circles in specific parts of the screen). We will create a webapp so course creators can "configure" checks for their assignments by combining these modules using drag and drop of these modules, without actually writing any code. Doing this will expand the usage of Processing to teaching coding with scalable grading. 

\section{Conclusion}
To achieve the evaluation aspect of the SuaCode course, assignments are inevitable. Since one of the goals of the course was to reach a large number of Africans, an efficient method needed to be introduced for evaluating a large number of students without needing to equally scale up the number of facilitators. We developed AutoGrad as a way to aid the evaluation process by grading the assignments automatically. The method leveraged the Java reflections API and regular expression to grade the assignments via static and dynamic code analysis. The deployment of AutoGrad resulted in reduced effort and time needed to grade the assignments required to complete the SuaCode course. Future work would involve deploying the system in the cloud with containers and virtual machines, improving the feedback given by AutoGrad and also developing a webapp to easily create grading modules for other Processing smartphone-based courses.

\begin{acks}
We are grateful to the Processing Foundation for supporting this work via the Processing Foundation Fellowship program. Also, we thank the AutoGrad team that contributed to developing the software. 
\end{acks}

\balance
\bibliographystyle{ACM-Reference-Format}
\bibliography{ref}

\end{document}